\documentclass[onecolumn,11pt]{IEEEtran}

\usepackage{times}
\usepackage{times,amssymb,amsmath,amsfonts,float,nicefrac,color,cite,bbm,mathrsfs,float,stmaryrd}
\usepackage{amsfonts}
\usepackage{latexsym}
\usepackage{amssymb}
\usepackage{theorem}
\usepackage{graphicx}
\usepackage{subcaption}
\usepackage{enumitem}
\usepackage{algorithm,algpseudocode}

\usepackage[top=1.3in,bottom=1.3in,left=1.25in,right=1.25in]{geometry}

\newtheorem{theorem}{Theorem}
\newtheorem{lemma}{Lemma}
\newtheorem{claim}{Claim}

\newtheorem{corollary}{Corollary}
\newtheorem{definition}{Definition}

\newtheorem{example}{Example}
\newtheorem{remark}{Remark}
\newtheorem{construction}{Construction}

\renewcommand{\thefigure}{\thesection.\arabic{figure}}
\usepackage{amsmath}

\allowdisplaybreaks

\makeatletter
\renewcommand{\@endtheorem}{\endtrivlist}
\makeatother

\makeatletter
\renewcommand{\thefigure}{{\@arabic\c@figure}}
\renewcommand{\fnum@figure}{{\bf Figure\,\thefigure}}
\makeatother

\newcommand\nc\newcommand
\nc{\cA}{\mathcal{A}}\nc{\cB}{\mathcal{B}}\nc{\cC}{\mathcal{C}}\nc{\cD}{\mathcal{D}}
\nc{\cE}{\mathcal{E}}\nc{\cF}{\mathcal{F}}\nc{\cG}{\mathcal{G}}\nc{\cH}{\mathcal{H}}
\nc{\cI}{\mathcal{I}}\nc{\cJ}{\mathcal{J}}\nc{\cK}{\mathcal{K}}\nc{\cL}{\mathcal{L}}
\nc{\cM}{\mathcal{M}}\nc{\cN}{\mathcal{N}}\nc{\cO}{\mathcal{O}}\nc{\cP}{\mathcal{P}}
\nc{\cQ}{\mathcal{Q}}\nc{\cR}{\mathcal{R}}\nc{\cS}{\mathcal{S}}\nc{\cT}{\mathcal{T}}
\nc{\cU}{\mathcal{U}}\nc{\cV}{\mathcal{V}}\nc{\cW}{\mathcal{W}}\nc{\cX}{\mathcal{X}}
\nc{\cY}{\mathcal{Y}}\nc{\cZ}{\mathcal{Z}}

\nc{\bba}{\mathbf{a}}\nc{\bbb}{\mathbf{b}}\nc{\bbc}{\mathbf{c}}\nc{\bbd}{\mathbf{d}}
\nc{\bbe}{\mathbf{e}}\nc{\bbf}{\mathbf{f}}\nc{\bbg}{\mathbf{g}}\nc{\bbh}{\mathbf{h}}
\nc{\bbi}{\mathbf{i}}\nc{\bbj}{\mathbf{j}}\nc{\bbk}{\mathbf{k}}\nc{\bbl}{\mathbf{l}}
\nc{\bbm}{\mathbf{m}}\nc{\bbn}{\mathbf{n}}\nc{\bbo}{\mathbf{o}}\nc{\bbp}{\mathbf{p}}
\nc{\bbq}{\mathbf{q}}\nc{\bbr}{\mathbf{r}}\nc{\bbs}{\boldsymbol{S}}\nc{\bbt}{\mathbf{t}}
\nc{\bbu}{\mathbf{u}}\nc{\bbv}{\mathbf{v}}\nc{\bbw}{\mathbf{w}}\nc{\bbx}{\mathbf{x}}
\nc{\bby}{\mathbf{y}}\nc{\bbz}{\mathbf{z}}

\nc{\sA}{\mathsf{A}}\nc{\sB}{\mathsf{B}}\nc{\sC}{\mathsf{C}}\nc{\sD}{\mathsf{D}}
\nc{\sE}{\mathsf{E}}\nc{\sF}{\mathsf{F}}\nc{\sG}{\mathsf{G}}\nc{\sH}{\mathsf{H}}
\nc{\sI}{\mathsf{I}}\nc{\sJ}{\mathsf{J}}\nc{\sK}{\mathsf{K}}\nc{\sL}{\mathsf{L}}
\nc{\sM}{\mathsf{M}}\nc{\sN}{\mathsf{N}}\nc{\sO}{\mathsf{O}}\nc{\sP}{\mathsf{P}}
\nc{\sQ}{\mathsf{Q}}\nc{\sR}{\mathsf{R}}\nc{\sS}{\mathsf{S}}\nc{\sT}{\mathsf{T}}
\nc{\sU}{\mathsf{U}}\nc{\sV}{\mathsf{V}}\nc{\sW}{\mathsf{W}}\nc{\sX}{\mathsf{X}}
\nc{\sY}{\mathsf{Y}}\nc{\sZ}{\mathsf{Z}}

\newcommand{\s}[1]{\mathsf{#1}}

\newcommand{\mathset}[1]{\left\{#1\right\}}

\newcommand{\parenv}[1]{\left( #1 \right)}
\newcommand{\sparenv}[1]{\left[ #1 \right]}

\newcommand{\bal}[1]{\begin{align}\label{#1}}
\newcommand{\eal}{\end{align}}

\newcommand{\pp}[1]{\left[#1\right]}

\newcommand{\abs}[1]{\left|#1\right|}
\newcommand{\pr}{\mathbb{P}}

\renewcommand{\leq}{\leqslant}

\renewcommand{\geq}{\geqslant}

\renewcommand{\Bbb}{\mathbb}

\newcommand{\Cref}[1]{Co\-ro\-lla\-ry\,\ref{#1}}

\renewcommand{\Bbb}{\mathbb}

\newcommand{\N}{{\Bbb N}}

\newcommand{\R}{{\Bbb R}}
\newcommand{\Z}{{\Bbb Z}}
\newcommand{\E}{{\Bbb E}}

\newcommand{\SigmaS}{\Sigma^{\star}}
\newcommand{\sRS}{\sR^{\star}}

\DeclareMathOperator{\ccap}{cap}

\newcommand{\0}{\mathbf{0}}

\newcommand{\1}{\mathbf{1}}

\newcounter{casenum}

\begin{document}

\title{Optimal Reference for DNA Synthesis}

\author{Ohad~Elishco\thanks{O.Elishco is with the Department of Electrical and Computer Engineering at Ben-Gurion university, {B}eer {S}heva, Israel (e-mail:  \texttt{elishco@gmail.com})}~~~~~~~~~~~~Wasim~Huleihel\thanks{W. Huleihel is with the Department of Electrical Engineering at Tel-Aviv university, {T}el {A}viv 6997801, Israel (e-mail:  \texttt{wasimh@tauex.tau.ac.il}).}}

\maketitle

\begin{abstract}

In the recent years, DNA has emerged as a potentially viable storage technology. DNA synthesis, which refers to the task of writing the data into DNA, is perhaps the most costly part of existing storage systems. Accordingly, this high cost and low throughput limits the practical use in available DNA synthesis technologies. It has been found that the \emph{homopolymer run} (i.e., the repetition of the same nucleotide) is a major factor affecting the synthesis and sequencing errors. Quite recently, \cite{MakRacRasYek2021} studied the role of batch optimization in reducing the cost of large scale DNA synthesis, for a given pool $\cal{S}$ of random quaternary strings of fixed length. Among other things, it was shown that the asymptotic cost savings of batch optimization are significantly greater when the strings in $\cal{S}$ contain repeats of the same character (homopolymer run of length one), as compared to the case where strings are unconstrained. 

Following the lead of \cite{MakRacRasYek2021}, in this paper, we take a step forward towards the theoretical understanding of DNA synthesis, 
and study the  homopolymer run of length $k\geq1$. 
Specifically, we are given a set of DNA strands $\cal{S}$, randomly drawn from a natural Markovian distribution modeling a general homopolymer 
run length constraint, that we wish to synthesize. For this problem, we prove that for any $k\geq 1$, the optimal reference strand, 
minimizing the cost of DNA synthesis is, perhaps surprisingly, the periodic sequence $\overline{\s{ACGT}}$. It turns out that tackling the homopolymer constraint of length $k\geq2$ is a challenging problem; our main technical contribution is the 
representation of the DNA synthesis process as a certain constrained system, for which string techniques can be applied.

\end{abstract}

\section{Introduction}\label{sec:Intro}

DNA data storage refers to the process of encoding (decoding) data to (from) synthesized sequences (or, strands) of DNA. 
Recently, there has been a growing interest in the problem of storing data in synthetic DNA molecules. 
Indeed, DNA, as a storage medium, has an enormous potential because of its high storage density compared to other conventional storage media. 
Unfortunately, however, the practical use of DNA as a provable efficient storage technology is currently sharply circumscribed mainly because of 
its high cost and very slow read and write duration. 

The typical approach used for producing DNA molecules is array-based DNA synthesis (see, e.g., \cite{Kosuri2014LargescaleDN}). 
In a nutshell, in this technique a machine synthesises a large number of DNA strands in parallel (referred to as information sequences), 
where each such strand is grown by one DNA character at each time step of the process. 
To that end, the machine generates multiple copies of a specific nucleotide, and these nucleotides are concatenated to a selected subset of 
the information sequences. 
The nucleotide that the machine generates at any given time is determined according to a {fixed} \emph{reference strand (or sequence)}. 
Specifically, as the synthesizer goes through this reference strand, the next character it reads in the reference strand is generated and 
concatenated to the selected subset of information sequences.  
This process terminates when the machine arrives at the end of the reference strand. 
It is evident that in order for the synthesis process to work, the reference strand must be a supersequence of all the information sequences.
This way each synthesized DNA strand is a subsequence of the reference strand and is synthesized. 
Accordingly, the length of the reference strand determines the \emph{synthesis time} of this DNA synthesis processes. In this paper, we will refer to the synthesis time as the \emph{cost}.

The encoding process in DNA data storage generates a list of DNA strands that need to be synthesized, by translating binary sources into 
strands of nucleotides (for example, by mapping two binary source bits into a single nucleotide). 
It is well-known that repetitions of the same nucleotide, also known as, a \emph{homopolymer run}, may significantly increase the chance of 
sequencing errors \cite{Bornholt16,Ross2012CharacterizingAM}. 
For example, it was observed in \cite{Ross2012CharacterizingAM} that a long homopolymer run (e.g. more than 4 nucleotides) results in a significant 
increase of insertion and deletion errors, and as so such long runs should be avoided. 
Therefore, encoding algorithms often ensure that the resulting strands do not contain long runs of the same character \cite{Goldman13,Organick18}.

\subsection{Main Conceptual and Technical Contributions}
In this paper, we consider the following meta generative model: 
we are given a set of DNA strands $\mathcal{S}_k$, drawn at random from a ``natural" distribution. 
This natural distribution aims to capture a general homopolymer run length constraint of length $k\geq 1$, i.e., the strands to synthesized 
are not allowed to contain $k+1$ repeated nucleotides. 
Our main goal is to find the optimal reference strand, where optimality is measured in terms of the synthesis cost, denoted by $\s{cost}(\mathcal{S}_k)$. This cost is defined as the length of the shortest common supersequence of all strands. 
To that end, we start by representing our DNA synthesis problem as a homopolymer run length constrained system associated with a unique, 
entropy maximizing Markov measure. 
This Markov measure plays the role of the previously mentioned ``natural'' distribution. 
Using a characterization of this measure we prove that for any $k\geq 1$, the optimal reference strand is $\overline{\s{ACGT}}$.\footnote{Given a string $w$, we denote by $\overline{w}\triangleq\ldots www \ldots$ the infinite sequence generated by repeated concatenations of $w$ with itself.}
Our analysis is a generalization of the single batch analysis for the special case of $k=1$ considered in \cite{MakRacRasYek2021}. 
This generalization to homopolymer run length constraint of length $k$ adds another level of difficulty that requires techniques from the field of constrained systems. 

\subsection{Related Work}

Most closely related paper to our work is \cite{MakRacRasYek2021}. In this paper, the authors study the role of batch optimization in reducing the cost of large scale DNA synthesis. They consider the cases where the strands to synthesised are either unconstrained or constrained, in the sense that the strands do not contain repeats of the \emph{same} character (homopolymers). Our paper generalize their results for the case of a general homopolymer run constraint of length $k\geq1$ (i.e., $k+1$ repeated characters are not allowed).

Similarly to \cite{MakRacRasYek2021}, our work is motivated by both theoretical and experimental papers that tackle the problem of reducing the cost of DNA synthesis. Specifically, in terms of theoretical results, a few recent works have considered coding-based approaches for the analysis of the cost. For example, in \cite{Lenz20}, it was shown that, for array-based DNA synthesis techniques, by introducing redundancy to the synthesized strands, one
can significantly decrease the number of synthesis cycles. The authors also derive the maximum amount of information per synthesis cycle
assuming that the strands to be synthesized is an arbitrary periodic sequence. In \cite{Anavy2019DataSI,Jain20,Lee2019TerminatorfreeTE}, a somewhat different synthesis model which assumes that information is stored based on run length patterns in the strings was considered, for which the amount of information bits per synthesis time unit is analyzed. Another large body of related work is on the study of the longest common subsequence (LCS) of random strings, e.g., \cite{Chvtal1975LongestCS,Danck1995UpperBF,Kiwi2004ExpectedLO,Navarro2001AGT,Lueker2009ImprovedBO,Bukh2019PeriodicWC,Christian16}. Specifically, it well-known that for two strings of length $n$, generated at random, the expected length of LCS is approximately $\gamma n$, where $\gamma>0$ is the Chv\'{a}tal-Sankoff constant. 

There is a large amount of prior works from the experimental point of view of DNA synthesis cost, e.g., \cite{Hannenhalli2002CombinatorialAF,Kahng2002BorderLM,Kahng2004ScalableHF,Rahmann2003TheSC,Ning2006TheDA,Rahmann2006SubsequenceCA,Kumar2010DNAMP,Ning2011TheMS,Srinivasan2014ARO,Hubbell1999FidelityPF,Colbourn2002ConstructionOO}, and many reference therein. 
The majority of these papers analyze empirically the cost when using $\overline{\s{ACGT}}$ as the reference strand. For example, in \cite{Rahmann2003TheSC}, it was observed that the cost of uniformly random strings is approximately Gaussian. In terms of the selection of a short reference strand, many algorithms have been proposed and tested empirically. Unfortunately, these heuristics do not provide any provable guarantees. 

\subsection{Organization}

The rest of this paper is organized as follows. In Section~\ref{sec:Setup} we formulate our model, state our main goals, and present our main findings. 
Section~\ref{sec:Proofs} is devoted to the proofs of our main results, and finally, in Section~\ref{sec:ConcandOut} we conclude our paper and present a few intriguing questions for future research.


\section{Setup and Problem Statement}\label{sec:Setup}

\begin{figure}[ht!]
    \centering
    \includegraphics{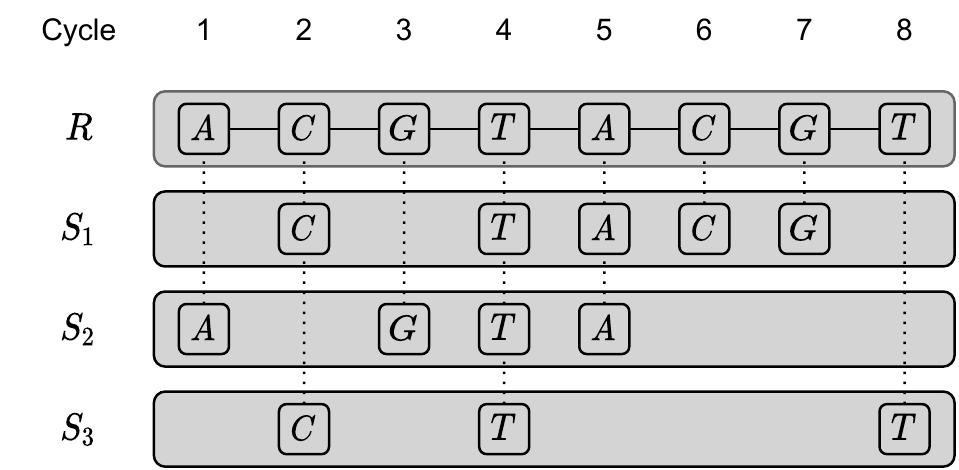}
    \caption{Synthesis of three strands $\s{S}_1 = (\s{CTACG}), \s{S}_2 = (\s{AGTA})$, and $\s{S}_3 = (\s{CTT})$ using the reference sequence $\s{R}=(\s{ACGTACGT})$. The strand $\s{S}_1$ is synthesized by attaching the nucleotides in cycles 2, 4, 5, 6, 7, $\s{S}_2$ is synthesized in cycles 1, 3, 4, 5, and similarly $\s{S}_3$ is synthesized in cycles 2, 4, 8. Henceforth, the cost of $\mathcal{S}=\{\s{S}_1,\s{S}_2,\s{S}_3\}$ is 8.}
    \label{fig:array-based}
\end{figure}

As mentioned above, the underlying problem in DNA synthesis is that strands of nucleotides with long repetitions of the same nucleotide are prone to errors, 
and thus we would like to avoid those DNA sequences with more than a fixed number $k\geq 1$ of consecutive nucleotides of the same type. \emph{In practice, the maximum run length of each symbol in each strand is at most three}. 
Throughout this paper, $n\in\N$ denotes the strand length. Consider the following definition.
\begin{definition}[Strands without $k$-homopolymers]
Fix $n,k\in\mathbb{N}$. 
Let $\mathcal{H}_{n,k}\subset\{\s{A},\s{C},\s{G},\s{T}\}^n$ be the subset of all strands of length $n$ with no $k+1$ consecutively repeated characters.
\end{definition}
Given $\mathcal{H}_{n,k}$, let $\mathcal{S}_k$ be a subset of strands in $\mathcal{H}_{n,k}$ with $|\mathcal{S}_k| = \s{M}$, for some $\s{M}\in\mathbb{N}$. The set $\mathcal{S}_k$ is the pool of strands to be synthesized. We consider a popular synthesis process that produces many strands in parallel, step-by-step, using a fixed supersequence denoted by $\s{R}\in\{\s{A},\s{C},\s{G},\s{T}\}^n$. We will refer to $\s{R}$ as the \emph{reference sequence}. The machine iterates through this supersequence one nucleotide at a time, and in each cycle, adds the next nucleotide to a subset of the strands. An example of this synthesis process is shown in Fig.~\ref{fig:array-based}.
Suppose we want to synthesize a set of DNA strands $\mathcal{S}_k$ using a reference strand $\s{R}$. Throughout the paper, we denote the length of the prefix of $\s{R}$ which we use for synthesis by $\s{cost}_{\s{R}}(\mathcal{S}_k)$. Our goal is to investigate the \emph{optimal} cost of synthesizing $\mathcal{S}_k$, defined formally as follows.
\begin{definition}[Cost of DNA synthesis]
The cost of synthesising $\mathcal{S}_k$, denoted by $\s{cost}(\mathcal{S}_k)$, is the length of the shortest common supersequence of all strands in $\mathcal{S}_k$. The shortest common supersequence, denoted by $\s{R}^\star\in\{\s{A},\s{C},\s{G},\s{T}\}^n$, is referred to as the optimal reference sequence, and consequently, $\s{cost}(\mathcal{S}_k) = \s{cost}_{\s{R}^\star}(\mathcal{S}_k)$.
\end{definition}
In \cite{MakRacRasYek2021} it was assumed that the strands in $\mathcal{S}_k$ are selected i.i.d. from $\mathcal{H}_{n,k}$ uniformly at random. Assuming a generative model for the selection of the strands in $\mathcal{S}_k$ is, in fact, quite natural. Indeed, a common practice in the encoding process of digital data in DNA is to randomize the input using a seeded pseudorandom number generator or
compressed and encrypted \cite{Chandak20}. Roughly speaking, this is done in order to reduce the frequency of undesirable patterns that may occur in strands that are used to represent the data.\footnote{Ensuring that strands look random also facilitates certain tasks that may be a part of the decoding process such as clustering, e.g., \cite{Organick18,NIPS2017_ab731488}, and trace reconstruction, e.g., \cite{Batu04,Holenstein08,Viswanathan08,Peres17,Peres18}.} Accordingly, following \cite{MakRacRasYek2021} we assume a certain generative model for the stands selection as well. In principle, we could assume the same uniformity assumption as in \cite{MakRacRasYek2021}, and analyze the cost of DNA synthesis. In this paper, however, we opted to focus on the probabilistic model below, which we found much more natural. To present our model, we start with a brief background on \emph{constrained systems}. At this point, we would like to mention that although in the above we have focused our attention on quaternary alphabet (motivated by DNA genetic codes), our results hold for any alphabet $\Sigma$ of cardinality $|\Sigma|=r\geq2$; henceforth, we shall follow this generality.

\subsection{Constrained Systems Recap}
We provide here a brief background on the topic of constrained system. 
The notations and definitions that we use throughout the paper follow \cite{MarRotSie2001}. 
For a natural number $n\in\N$, we denote by $[n]$ the set $[n]=\mathset{0,1,\dots,n-1}$, and for a number $t\in\mathbb{N}$ we 
let $t+[n]=\mathset{t+j : j\in [n]}$. 
Fix a finite alphabet $\Sigma$ of size $|\Sigma|=r$. 
We denote by $\Sigma^\star$ the set of all finite words over $\Sigma$ and for $w\in\Sigma^\star$ we denote by $|w|$ its length. 
For $w,u\in\Sigma^\star$ we denote by $wu$ the word obtained by concatenating $u$ to $w$ and for $n\in \N$, $w^n$ denotes the concatenation of $w$ 
with itself $n$ times.
For a word $w=(w_0,w_1,\dots,w_{n-1})\in\Sigma^n$ and for a set $\cI\subseteq [n]$ we denote by $w_{\cI}$ the word obtained by 
restricting $w$ to the coordinates in $\cI$. 
For example, if $w=(w_0,\dots,w_{n-1})$ with $n\geq 4$ and $\cI=1+[3]=\mathset{1,2,3}$, then $w_{\cI}=(w_1,w_2,w_3)$.

A constrained system is defined by a (possibly infinite) set $\cF$ of finite words, $\cF\subseteq \SigmaS$. We think of the set $\cF$ as a set of forbidden patterns. A constrained system $\cS=\cS_{\cF}$ comprises of the set of all finite words that do not contain any word from $\cF$ as a subword, i.e., $w\in \cS_{\cF}$ if there are no pairs of indices $i,j\in\N$, $i<j$ and for which $w_i w_{i+1}\dots w_j=\alpha$, for some $\alpha \in \cF$. 

An equivalent way to describe a constrained system is using a graph. Specifically, let $G=(V,E,L)$ be a finite graph with $V$ being its vertex set, $E\subseteq V\times V$ a set of (directed) edges, and $L:E\to\Sigma$ a label function. 
A path $\gamma$ of length $n$ in $G$ is a sequence of edges $\gamma=(e_0,e_1,\dots,e_{n-1})\in E^n$ where $e_i=(v_i,v_{i+1})$ (notice that $e_i$ ends in the vertex $e_{i+1}$ starts from). The label of $\gamma$ is the word $L(\gamma)\triangleq L(e_0)L(e_1)\dots L(e_{n-1})$ and we say that $\gamma$ starts at $v_0$ and ends at $v_n$. 
A constrained system $\cS$ is the set of all finite words obtained from reading the labels of paths in a labeled graph $G$. 
We say that $G$ is a presentation of $\cS$, or $G$ presents $\cS$. 
Notice that there are many other different presentations for the same system.

A simple description of a labeled graph $G$ can be obtained using the adjacency matrix $A=A_G$. 
The adjacency matrix is a $|V|\times |V|$ matrix where the $(i,j)$ entry is the number of edges going from state $v_i$ to state $v_j$ in $G$. Fig.~\ref{fig:cons_sys0} below illustrates a graph that presents a system $\cS$ that comprises of all the words in which no symbol appears next to itself.
\begin{figure}[ht!]
    \centering
    \includegraphics{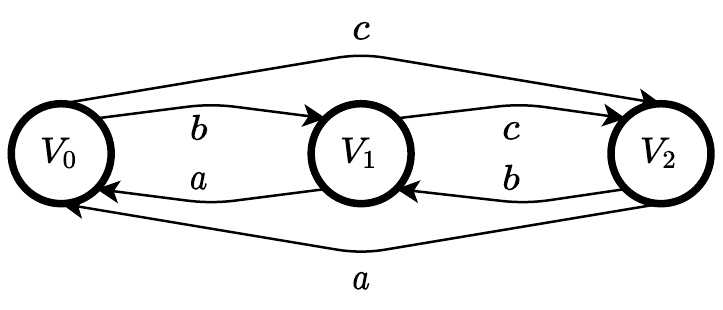}
    \caption{A presentation of a system $\cS$ that comprises of all the words in which no symbol appears next to itself.}
    \label{fig:cons_sys0}
\end{figure}
A useful property of constrained systems is \emph{irreducibility}. A constrained system $\cS$ is called 
irreducible if for every $\alpha,\beta\in \cS$ there is a word $\gamma\in \cS$ such that $\alpha\gamma\beta\in \cS$. 
An equivalent requirement for $\cS$ to be irreducible is the existence of a presentation $G$ of $\cS$ in which there is a path between every two vertices (the graph $G$ is strongly connected). 

The \emph{capacity} of a constrained system $\cS$ is, intuitively speaking, a measure for the complexity of the system. 
It is defined as $\ccap(\cS)\triangleq\lim_{n\to\infty} \frac{1}{n}\log |\cB_n(\cS)|$, where $\cB_n(\cS)\triangleq\mathset{s\in \Sigma^n: s\in \cS}$ is 
the set of all $n$-length words in $\cS$ and the logarithm is to the base of $|\Sigma|=r$. 
It is a well known fact that the limit in this definition exists \cite{MarRotSie2001}. 
If $G$ is a graph that presents an irreducible constrained system $\cS$, the Perron-Frobenius Theorem provides a characterization of $\ccap(\cS)$ using the eigenvectors and eigenvalues of the adjacency matrix $A_G$ (see, for example \cite[Ch. 3]{MarRotSie2001}). 
\begin{lemma}[Perron-Frobenius Theorem]
\label{th:perron_frobenius}
Let $\cS$ be an irreducible constrained system presented by a strongly connected graph $G$. Then the following hold.
\begin{enumerate}
    \item The adjacency matrix $A_G$ has a real, simple, maximal eigenvalue $\lambda$ called the Perron eigenvalue. 
    \item $A_G$ has a positive left eigenvector $\xi$ and a positive right eigenvector $\phi$ associated with $\lambda$, called Perron eigenvectors.
    \item The only eigenvectors with all positive components are the Perron eigenvectors.
\end{enumerate}
\end{lemma}
It can be shown that $\ccap(\cS)=\log \lambda$, where $\lambda$ is the Perron eigenvalue of $A_G$ \cite[Ch. 3]{MarRotSie2001}.

\subsection{Probabilistic Viewpoint of Constrained Systems}
To relate our DNA synthesis problem to constrained systems we need to associate $\cS_k$ with a Markov chain. 
Let $\cS$ be an irreducible constrained system over the alphabet $\Sigma=[r]$ for some $r\in \N$. 
Let $G=(V,E,L)$ be a graph that presents $\cS$ with its corresponding adjacency matrix $A=A_G$. 
There is a unique (stationary) Markov measure $\mu$, such that the $n$ marginal of $\mu$ is supported on $\cB_n(\cS)$ for $n\in \N$, and with 
Shannon-entropy rate equals to the capacity $\ccap(\cS)=\log \lambda$ (see \cite[Ch. 3]{MarRotSie2001}, \cite[Ch. 4]{LinMar85} or \cite[Ch. 8]{Wal1982}). 

The measure $\mu$ can be described by a stationary distribution $\pi$ over the set of vertices $V$ and a transition probabilities matrix $Q\in \R^{|V|\times |V|}$. Given $\pi,Q$, the probability of a path $\gamma=(e_0,\dots,e_{n-1})$ in $G$ is given by 
\begin{align}
\pr(\gamma)=\pi(v_0) Q_{v_0,v_1} Q_{v_1,v_2}\dots Q_{v_{n-1},v_n},
\end{align}
where $e_i=(v_i,v_{i+1})$. Since there are several paths with the same label, the probability of a word $w$ under $\mu$ is $\sum_{\gamma : L(\gamma)=w} \pr(\gamma)$. The constructions of $\pi$ and $Q$ are well-known \cite{MarRotSie2001}. Specifically, let $\xi,\phi$ be the left and right Perron eigenvectors of $A_G$ associated with the Perron eigenvalue $\lambda$, normalized such that $\sum_{i\in V}\xi_i \phi_i=1$. 
Then, the transition matrix $Q$ is given by $Q_{i,j}=\frac{(A_G)_{i,j} \phi_j}{\lambda \phi_i}$, and the stationary distribution is given by $\pi_i=\phi_i \xi_i$. Finally, the entropy rate of this Markov chain is exactly $\ccap(\cS)$, and $\mu$ is the unique (stationary) measure that maximizes the entropy ($\mu$ is a Markov measure of $[r]^{\Z}$ obtained using Kolmogorov's extension theorem).

\subsection{Problem Statement and Main Result}

We now state our problem and main result rigorously. 
First, we define the relevant constrained system we study in this paper, 
dubbed $k$-run length constraints.
\begin{definition}
Let $\Sigma$ be a finite alphabet and fix $k\in\N$. Denote by $\cF_k$ the set of all words of length $k+1$ that comprise of a single symbol 
$\cF_k\triangleq\mathset{a^{k+1} : a\in\Sigma}$. 
The $k$-run length (constrained) system is denoted by $\cS_k^{\Sigma}$ and is defined by the set $\cF_k$, $\cS^{\Sigma}_k=\cS_{\cF_k}$. 
In words, $\cS_k^{\Sigma}$ is the set of all finite words over $\Sigma$ in which there are no runs of length $k+1$. 
When $\Sigma$ is clear from the context we will write $\cS_k$ instead of $\cS_k^{\Sigma}$.
\end{definition}

\begin{example}
Let $\Sigma=[2]=\mathset{0,1}$ and let $\cS_2^{\Sigma}$ be the $2$-run length constrained system. 
The system $\cS_2^{\Sigma}$ comprises of all the finite binary words $w$ such that $w$ does not contain any of the patterns $000,111$, i.e., 
there are no triples of similar consecutive symbols. 
\end{example}

Fix an alphabet $\Sigma$ and numbers $k,n,\sM\in\N$. 
We are to synthesize $\sM$ sequences over the alphabet $\Sigma$, where each sequence is chosen independently at random from $\cB_n\parenv{\cS_k^{\Sigma}}$, according to the (unique, stationary) entropy maximizing measure $\mu$. 
The multiset $\cS_k$ of $\sM$ strands is called a batch. 
Our primary goal is to characterize the optimal reference strand for synthesizing all the strands in a batch $\cS_k$. 
Our main result is the following.
\begin{theorem}[Optimal reference]
\label{th:main}
For an alphabet $\Sigma=[r]$ and for any $k\geq 1$, let $\cS_k\subseteq \cB_n\parenv{\cS^{[r]}_k}$ be a batch of $\sM$ i.i.d. sequences chosen according to $\mu$. 
Then, with probability at least $1-1/n$, there exists a constant $\s{C}>0$, such that for any $\epsilon>0$,
    $$[\s{C}-\epsilon]\cdot n\leq\s{cost}_{\sRS}(\mathcal{S}_k)\leq\s{cost}(\mathcal{S}_k)\leq[\s{C}+\epsilon]\cdot n,$$
    where $\sRS=\overline{012\dots (r-1)}$.
\end{theorem}
While in the practice of DNA synthesis the parameters $n$, $\s{M}$, and $k$ are concrete numbers, to facilitate the asymptotic study of the problem we focus on the following relevant scenario: $n$ is growing, $\s{M}$ is significantly larger than but polynomial in $n$, and $k$ is either a constant or a growing function of $n$.

\section{Proof of Theorem~\ref{th:main}}\label{sec:Proofs}
In this section we prove Theorem \ref{th:main}. We start by showing that for every $r\in\N$ and every $k\in \N$, $\cS^{\Sigma}_k$ is an irreducible constrained system. 
To that end, we construct an irreducible graph presentation of $\cS_k$ which is based on the de Bruijn graph of span $k$. 
We also show that the adjacency matrix $A_G^{\Sigma}$ of this graph $G$ can be obtained from the adjacency matrix of the de Bruijn graph of span $k$. 
We then use the structure of the adjacency matrix $A_G^{\Sigma}$ to find the right Perron eigenvector of the matrix. 
By relating the constrained system to Markov chains, we use the right Perron eigenvector to describe the transition probabilities of the chain. 
This description allows us to pin-point a useful characteristic of the transition probabilities, which is then used, together with Hoeffding's inequality, to 
prove Theorem \ref{th:main}.

\subsection{Constructing $G$ and $A_G$ from de Bruijn graph} \label{suba}
We begin by constructing the de Bruijn graph $G'$ of span $k$. 
Let $G'=G'([r],k)=(V',E',L')$ where $V'=[r]^k$ is the set of all $k$-length words over $[r]$. 
The set $V'$ inherits the natural order obtained by interpreting the vertices as numbers written in their base-$r$ representation. 
To construct the set of edges $E'$, draw a directed edge from $u=(a_0,\dots,a_{k-1})\in V'$ to $v=(b_0,\dots,b_{k-1})\in V'$ 
if $b_i=a_{i+1}$ for every $i\in [k-1]$. In words, we draw an edge from $u$ to $v$ if the $k-1$ suffix of $u$ is equal to the $k-1$ prefix of $v$. 
The label of the edge $(u,v)$ is $L'((u,v))=b_{k-1}$. 
The graph $G'$ is called the \emph{de Bruijn graph of span $k$}. 

Notice that every path of length $k$ in $G'$ that ends at a vertex $v=(b_0,\dots,b_{k-1})$, yields the word $L(\gamma)=b_0\dots b_{k-1}$. 
Thus, with a slight abuse of notation we will sometimes use the vertex symbol $v$ instead of its corresponding $k$-tuple $b_0\dots b_{k-1}$.
Notice that an appearance of an edge $(u,v)$ in a path in $G'$ can be related to the appearance of the $k+1$-length subword 
$u L((u,v))=u b_{k-1}=a_0 v$. 

The graph that presents $\cS_k$ can be easily obtained from the de Bruijn graph $G'$ by removing some edges. 
Since runs of length $k+1$ are forbidden in $\cS_k$, to obtain a presentation $G=G(\Sigma,k)=(V,E,L)$ of $\cS_k$ we let $V=V'$, and $E=E'$ after 
eliminating self loops from $E'$, i.e., we take $E=E'\setminus \mathset{(v,v) : v\in V}$. 
Notice that the only self loops in $E'$ are edges $(v,v)\in E'$ of the form $v=a^k$ where $a\in \Sigma$, i.e., $v$ corresponds to a $k$-length word comprises of a single letter. 
Since self loops in $G'$ correspond to $k+1$-tuples of the form $a^{k+1}$ for some $a\in\Sigma$, and since $G$ is obtained after all the self-loops are removed from $G'$, the graph $G$ presents the system $\cS_k$.
From the structure of $G$ we immediately obtain that there is a path from any vertex $u$ to any vertex $v=(b_0,\dots,b_{k-1})$ in $G$, by walking over the edges labeled with $b_0,b_1,\dots,b_{k-1}$. The only case in which this is not possible is 
when $uv$ contains a forbidden pattern. In this case, the path $c b_0,\dots,b_{k-1}$, for some $c\neq b_0$, is a path from $u$ to $v$ in $G$. 
This immediately implies the following corollary.
\begin{corollary}
\label{cor:S_k_irreducible}
The $k$-run length system $\cS_k^{\Sigma}$ is irreducible.
\end{corollary}

\begin{remark}
\label{rem:lossless}
Notice that the presentation of $\cS_k$ described above is not the minimal (in terms of number of vertices) representation. We will, however, consider this presentation throughout the paper since it is more accessible for analysis. This accessibility follows from two facts. The first, is that the adjacency matrix of $G$ has some structure which is preserved when $k$ is increased. The second, is that this presentation is lossless, which means that fixing a starting vertex and an ending vertex, different paths generate different words. The latter property will be used later to bound the cost.
\end{remark}

Similarly to the construction of $G$, the adjacency matrix $A_G$ of the graph $G$ can be obtained from the adjacency 
matrix of the de Bruijn graph of span $k$. Obviously, to get $A_G$ from $A_{G'}$ we only need to set zero the entries $A_{G'}(v,v)$. 
The following lemma provides a formal construction.
\begin{lemma}
\label{lem:adj_mat_of_S_k}
Let $G=G(\Sigma,k)$ be the graph presentation of $\cS_k$. Then, the adjacency matrix of $G$, denoted by $A^{\Sigma}_k\in \N^{r^k\times r^k}$, is given by 
\begin{align}
\label{eq:A_G}
    (A^{\Sigma}_k)_{i,j}= \begin{cases}
			0, & \text{if } j\notin \mathset{(i\bmod r)\cdot r + [r]} \;\text{or}\; i=j=0\pmod{\frac{r^k -1}{r-1}}\\
            1, & \text{otherwise}.
		 \end{cases}
\end{align}
We will write $A_k$ instead of $A_k^{\Sigma}$ when the alphabet is clear from the context or if it is irrelevant.
\end{lemma}

The structure of the adjacency matrices $A_k$ is best seen by examples.
\begin{example}
\label{ex:adj_mat}
For $\Sigma=[2]$, the adjacency matrices $A^{[2]}_2$ and $A^{[2]}_3$ corresponding to $\cS_2^{[2]}$ and $\cS_3^{[2]}$, respectively, are:
\[A^{[2]}_2=\begin{bmatrix} \mathbf{0}&1&0&0\\ 0&0&1&1 \\ 1&1&0&0 \\ 0&0&1&\mathbf{0} \end{bmatrix}, \; 
A^{[2]}_3=\begin{bmatrix} \mathbf{0}&1&0&0&0&0&0&0\\ 0&0&1&1&0&0&0&0\\0&0&0&0&1&1&0&0\\ 0&0&0&0&0&0&1&1\\ 1&1&0&0&0&0&0&0\\ 0&0&1&1&0&0&0&0\\0&0&0&0&1&1&0&0\\ 0&0&0&0&0&0&1&\mathbf{0} \end{bmatrix}.\] 
For $\Sigma=[3]$, the adjacency matrix $A^{[3]}_2$ corresponding to $\cS_2^{[3]}$ is  
\[A^{[3]}_2=\begin{bmatrix} \mathbf{0}&1&1&0&0&0&0&0&0\\ 0&0&0&1&1&1&0&0&0\\ 0&0&0&0&0&0&1&1&1\\ 1&1&1&0&0&0&0&0&0\\ 0&0&0&1&\mathbf{0}&1&0&0&0\\ 0&0&0&0&0&0&1&1&1\\ 1&1&1&0&0&0&0&0&0\\ 0&0&0&1&1&1&0&0&0\\ 0&0&0&0&0&0&1&1&\mathbf{0} \end{bmatrix}.\]
The numbers in bold are those numbers that were changed from the standard de Bruijn matrix due to the removal of self-loops.
Notice that $A_2^{[2]}$ comprises of $|\Sigma|=2$ ``main blocks" where each block comprises of $2^{2-1}$ ``stairs" of length $|\Sigma|=2$ each, and $A_3^{[2]}$ comprises of $|\Sigma|=2$ main blocks where each block comprises of $2^{3-1}$ ``stairs" of length $2$ each.
The matrix $A_2^{[3]}$ comprises of $3$ main blocks where each block comprises of $3^{2-1}$ ``stairs" of length $3$ each.
In general, if $\cS_k^{\Sigma}$ is a $k$-run length constrained system over $\Sigma=[r]$, then the corresponding adjacency matrix, $A_k^{[r]}$, comprises of 
$r$ main blocks, each block has $r^{k-1}$ ``stairs"  where each stairs contains $r$ ones or $r-1$ ones and a ``bold" zero.
Moreover, the zeros in bold correspond to vertices that have self loops in the de Bruijn graph of span $k$. Those vertices are exactly the vertices that 
are labeled as $a^k$ for some $a\in\Sigma$. Interpreting the labels of the vertices as base $r$ numbers, we obtain that for every $i\in [r]$, the $i$th number in the $i\frac{r^{k-1}-1}{r-1}$th ``stair" of the $i$th block is a (bold) $0$.
\end{example}

We notice that for the special case of $k=1$, the adjacency matrix $A_1$ has the following form. 
With $\Sigma=[r]$, observe that every row contains $r-1$ ones and a single zero. 
The zeros are located on the diagonal of $A_1$. 
It is immediate to see that the all-one vector $\1$ serves as a left and as a right eigenvector with eigenvalue $r-1$ (considered as a row or a column vector). 
Thus, normalizing the left and right eigenvectors, $\xi,\phi$ such that $\sum_{i\in [r]} \xi_i \phi_i=1$ we obtain that 
$\phi_i=\xi_i=\frac{1}{\sqrt{r}}$. As a result, the characterization of $\mu$ is simple.
\begin{lemma}
For the special case of $k=1$, the Markov chain $\mu$ is the uniform distribution. 
\end{lemma}
Therefore, in the special case of $k=1$, considered in \cite{MakRacRasYek2021}, we see that our generative assumption on the set of strands $\mathcal{S}_k$ coincides with the probabilistic model assumed in \cite{MakRacRasYek2021}.

\subsection{The Structure of $\mu$ Implies Theorem \ref{th:main}}
In this section we show that Theorem \ref{th:main} follows from the structure of $\mu$. 
Specifically, Theorem~\ref{th:main} follows almost immediately from the following result.
\begin{theorem}
\label{th:main2}
Let $\Sigma$ be a finite alphabet and let $k\in\N$. 
Let $\mu$ be the Markov measure that is associated with the system $\cS_k$. 
Then for every $a\neq b\in\Sigma$ and every $i$, $\mu(ab^i)> \mu(ab^{i+1})$. 
In words, the probability of seeing $i$ consecutive symbols is decreasing with $i$. 
Moreover, for every $a,b,c\in\Sigma$ with $a\neq b$ and $a\neq c$, 
\[\mu(ab^i)=\mu(ba^i)=\mu(ac^i),\] 
for every $i$.
\end{theorem}

Let us now show that Theorem \ref{th:main} follows from Theorem \ref{th:main2}. 
The proof contains two parts. At first, we show that $\sRS$ is optimal for a single sequence $\sS$, chosen according to $\mu$. 
This is done using a stochastic domination argument. 
We then apply Hoeffding's inequality for Markov chains to prove the optimality of $\sRS$ for a batch $\cS_k$ of $\sM$ sequences. 
The following result is needed (see \cite[Corollary 1]{Mou2020} or \cite{Moulos20}). 
\begin{lemma}[Hoeffding's inequality for Markov chains \cite{Mou2020}] 
\label{th:Hoeffding_Markov}
Let $\mathset{X_i}_i$ be an irreducible Markov chain on a finite state space $V$, with initial distribution $\pi$ (the stationary distribution). 
Let $f:V^2\to [a,b]$ be a real-valued function evaluated on the edges of the Markov chain. Then for any $t>0$, 
\begin{align}
    \pr_{\pi}\parenv{\abs{\frac{1}{n}\sum_{i=0}^{n-1} f(X_i,X_{i+1}) - \E_{\pi}\sparenv{f(X_0,X_1)}}\geq t}\leq 2e^{-\frac{2nt^2}{(b-a)^2 \s{HitT}^2}},\label{eq:Hoeffding_Markov}
\end{align}
where $\pr_{\pi}(\cdot),\E_{\pi}[\cdot]$ are the probability and expected value when $\pi$ is the initial distribution, and 
\[\s{HitT}\triangleq\max_{(x_0,x_1),(y_0,y_1)\in V^2} \E\sparenv{T_{(y_0,y_1)} ~|~ (X_0,X_1)=(x_0,x_1)},\] 
with $T_{(y_0,y_1)}\triangleq\inf\mathset{n\geq 0 |~ (X_{n+1},X_{n+2})=(y_0,y_1)}$ is the first hitting time of edge $(y_0,y_1)$.
\end{lemma}

We are now in a position to prove Theorem \ref{th:main} using Theorem \ref{th:main2}.
\begin{IEEEproof}[Proof of Theorem \ref{th:main}] 
Let $\sS\in \cS_k$ be a length-$n$ sequence chosen according to $\mu$ and let $\sR\in\SigmaS$ be any reference sequence. 
Extend $\sR$ by concatenating $\overline{01\dots (r-1)}$, so that $\sR$ will surely be a supersequence of $\sS$. 
Let $\tau_i(\sR)$ denote the index of the symbol of $\sR$ that is used to print the $i$th symbol of $\sS$. 
Define $X_0\triangleq\tau_0(\sR)$ and $X_i\triangleq\tau_i(\sR)-\tau_{i-1}(\sR)$, for $i\geq 1$. 
Notice that the probability that $\s{cost}(\sS)\leq \rho$, for some $\rho\in\N$, is given by 
\[\pr(\s{cost}(\sS)\leq \rho)=\pr\parenv{\tau_n(\sR)\leq \rho}=\pr\parenv{\sum_{i=0}^{n-1}X_i\leq \rho}.\] 
Now let $\sRS=\overline{01\dots(r-1)}$ and similarly, let $\tau_i(\sRS)$ denote the index of the symbol of $\sRS$ that is used to print 
the $i$th symbol of $\sS$. 
Define $Y_0\triangleq\tau_0(\sRS)$ and $Y_i\triangleq\tau_i(\sRS)-\tau_{i-1}(\sRS)$, for $i\geq 1$. For an arbitrary (predetermined and known) reference $\sR$, the random variables $\mathset{X_i}$ are not i.i.d. 
However, the support of these random variables contains at most $r$ integers, which are the distances to the next occurrences of the $r$ symbols. 
In some cases, the support contains at most $r-1$ integers due to the 
run length constraint. The support of the random variables $\mathset{Y_i}$ 
is $\mathset{1,2,\dots,r}$ 
(or, due to the run length constraint, $\mathset{1,2,\dots,r-1}$), i.e., 
the support comprises of the minimal integers possible. 
Then, using Theorem \ref{th:main2} we obtain that for $t\in\N$ and $l\in [n]$,
\begin{equation}
\label{eq:num_4}
\begin{split}
    \pr(\tau_l(\sR)\geq t)&= \pr\parenv{X_l+\tau_{l-1}(\sR)\geq t}\\ 
    &= \E\sparenv{\pr\parenv{X_l+\tau_{l-1}(\sR) \geq t ~|~ \tau_{l-1}(\sR)}}\\ 
    &\stackrel{(a)}{\geq}  \E\sparenv{\pr\parenv{Y_l+\tau_{l-1}(\sR) \geq t ~|~ \tau_{l-1}(\sR)}}\\
    &= \pr(Y_l+\tau_{l-1}(\sR) \geq t),
    \end{split}
\end{equation}
where $(a)$ follows from Theorem \ref{th:main2} since the probability that $S_l=S_{l-1}$ is smaller than the probability that $S_l\neq S_{l-1}$. 
Note that Theorem \ref{th:main2} applies here despite the conditioning on $\tau_{l-1}(\s{R})$ in \eqref{eq:num_4}. Indeed, conditioning on $\tau_{l-1}(\s{R})$ 
fixes the vertex at time $l-1$, or alternatively, fixes the symbol at time $l-1$. 
Next, for $j\leq l\in [n]$, define $\bar{X}_j^l\triangleq\sum_{i=j}^l X_i$ and $\bar{Y}_j^l\triangleq\sum_{i=j}^l Y_i$. Let $j<l<n$ and consider $\pr\parenv{\bar{X}_0^j+\bar{Y}_{j+1}^l\geq t}$. We have 
\begin{equation}
\label{eq:num_5}
\begin{split}
    \pr\parenv{\bar{X}_0^j+\bar{Y}_{j+1}^l\geq t}&= \pr\parenv{\bar{X}_0^{j-1}+X_j+\bar{Y}_{j+1}^l\geq t}\\ 
    &= \E\sparenv{\pr\parenv{\bar{X}_0^{j-1}+X_j+\bar{Y}_{j+1}^l\geq t ~|~ X_0^{j-1},Y_{j+1}^l}} \\  
    &\stackrel{(b)}{\geq}  \E\sparenv{\pr\parenv{\bar{X}_0^{j-1}+Y_j+\bar{Y}_{j+1}^l\geq t ~|~ X_0^{j-1},Y_{j+1}^l}} \\ 
    &= \pr\parenv{\bar{X}_0^{j-1}+\bar{Y}_j^l \geq t}. 
    \end{split}
\end{equation}
To account for $(b)$ notice that conditioning on $(X_0^{j-1},Y_{j+1}^l)$ is equivalent to using the reference sequence $\sR$ up to the printing of $\sS_j$ and then 
switching to the reference sequence $\sRS$, while knowing the first $j$ symbols $\sS_0^{j-1}$, and knowing the rest of the symbols $\sS_{j+1}^l$ as a function of $\sS_j$. 
We stress that although conditioning on $X_0^{j-1}$ is the same as conditioning on $\sS_0^{j-1}$ (since the reference 
sequence is known and $X_0$ implies $S_0$), conditioning on $Y_{j+1}^l$ is not the same as conditioning $S_{j+1}^l$ (since $S_j$ is unknown).
This is because $Y_i$ is the \emph{number of steps} it takes to print the $i$th symbol in $\sS$ after the $(i-1)$th symbol was printed, 
using the reference $\overline{01\dots(r-1)}$. 
Thus, $(b)$ follows from Theorem \ref{th:main2} together with the symmetry of the run length constraint. Indeed, given the prefix $\sS_0^{j-1}$, the probability that $\sS_j=a\neq \sS_{j-1}$ is the same for every $a\in\Sigma\setminus {\sS_{j-1}}$ and is larger than the probability that $\sS_j=\sS_{j-1}$. 

From symmetry, it is clear that the first symbol is distributed uniformly over $[r]$; hence $\pr(X_0\geq t)\geq \pr(Y_0\geq t)$. 
This, together with \eqref{eq:num_4} and \eqref{eq:num_5}, imply that for every $\sR$, 
\[\pr\parenv{\tau_n(\sR)\geq t}\geq \pr\parenv{\tau_n(\sRS)\geq t},\] 
or, equivalently, 
\begin{align}
\label{eq:Base_inequality}
\pr\parenv{\bar{X}_0^n\leq t }\leq \pr\parenv{\bar{Y}_0^n\leq t}.
\end{align}

To show that $\sRS$ is optimal for a batch $\cS_k$ of $\sM$ i.i.d. sequences, we use Lemma \ref{th:Hoeffding_Markov}. 
As mentioned in Remark \ref{rem:lossless}, the graph $G$ that presents $\cS_k$ is lossless, i.e., two different paths that start at the same vertex and end 
at the same vertex generate different words. Moreover, after $k$ steps, every path with the same $k$-length prefix arrive at the same vertex. 
Thus, the Markov process $\mathset{Y_i}_i$ can be obtained using the graph $G=(V,E,L)$ that presents the system $\cS_k$ by replacing the label function $L$ 
with $f:V^2\to [0,r]$, denoting the cost of synthesis for each edge. For example, with $r=k=3$, the value of $f(000,002)=2$ and $f(010,100)=3$. 
Under this ``new" setting, the strong law of large numbers for Markov chains implies that the Ces\`aro mean of $Y_i$s converges to $\E[f]$. 
We can now use Hoeffding's inequality as follows. 
Since the graph $G$ is irreducible, denoting $\rho\triangleq(\E[f]-\epsilon)n$ and using \eqref{eq:Hoeffding_Markov} yields 
\[\pr\parenv{\bar{Y}_0^{n-1}\leq m}\leq 2e^{-C\cdot n},\] 
where $C= \frac{2\epsilon^2}{r^2 \s{HitT}^2}>0$ is a (finite) constant due to irreducibility of $G$. 

Overall, we obtain 
\begin{align*}
    \pr\parenv{\s{cost}(\cS)\leq \rho}&\leq \pp{\pr\parenv{\s{cost}(\sS)\leq \rho}}^{\sM} \\
    &=\pp{\pr(\tau_n(\sS)\leq \rho)}^{\sM} \\
    &=\pp{\pr\parenv{\bar{X}_0^{n-1}\leq \rho}}^{\sM}\\
    &\leq \pp{\pr\parenv{\bar{Y}_0^{n-1}\leq \rho}}^{\sM}\\
    &\leq 2^{\sM}\cdot e^{-C\cdot \sM n}
\end{align*}
where the first inequality follows since $\cS$ is a set of $\sM$ sequences chosen in an i.i.d. fashion and $\sS$ is a single sequence.
Thus, the probability that the cost of synthesis is less than $n\E[f]$ goes to $0$ for large $n$. 
On the other hand, let $\rho'\triangleq(\E[f]+\epsilon)n$ and obtain 
\begin{align*}
    \pr\parenv{\s{cost}(\cS)\geq \rho'}&\leq \sM \cdot\pr\parenv{\s{cost}(\sS)\geq \rho'}\\
    &\stackrel{(a)}{\leq} \sM \cdot\pr\parenv{\bar{Y}_0^{n-1}\geq \rho'}\\
    &\stackrel{(b)}{\leq} \sM\cdot 2e^{-C\cdot n}\to 0.
\end{align*}
where $(a)$ follows by choosing a specific reference sequence $\sRS$ and $(b)$ follows from Lemma \ref{th:Hoeffding_Markov}. 
This concludes the proof.
\end{IEEEproof}

Thus, the rest of the paper is devoted for the proof of Theorem \ref{th:main2}. 

\subsection{Proof of Theorem \ref{th:main2}}
Throughout, when considering a graph presentation of $\cS^{[r]}_k$, we will always use the graph $G$ obtained from the de-Bruijn graph as explained 
in Section \ref{suba}. Therefore, the adjacency matrix of the graph presenting $\cS_k$, is
the adjacency matrix $A_k$ given in Lemma \ref{lem:adj_mat_of_S_k}. 

For an alphabet $[r]$, every vertex in $G$ can be described by its corresponding $k$-tuple, or alternatively, as a number in a base-$r$ representation. 
Thinking of vertices as numbers in their base-$r$ representation will make the proofs easier to follow. 
For example, the zeros in bold in $A_k^{[r]}$ in Example \ref{ex:adj_mat} correspond to vertices of the form $a^k$ for some $a\in \Sigma$. 
Thinking of vertices as a base-$r$ numbers, vertices of the form $a^k$ correspond to $i\frac{r^k-1}{r-1}$ for a number $i\in [r]$. 
This explains the row and column numbers in which a bold zero will appear. 

We start with an explicit formula for the capacity of $\cS_k^{[r]}$ (see, e.g., \cite{NguCaiImmKia2021,ImminkCap}).
\begin{theorem}[Capacity formula]
\label{th:cap_k_run_lim}
Let $\cS_k^{[r]}$ be a $k$-run length system ($k>0$) over the finite alphabet $\Sigma=[r]$. The capacity of $\cS_k^{[r]}$ is given by $\ccap(\cS_k)=\log \lambda_k$ where $\lambda_k$ is the largest real root of the polynomial 
\begin{align}
    \label{eq:def_polynom_h}
    h_{k,r}(z):=z^k-\sum_{i=0}^{k-1} (r-1)z^i=\frac{z^k(z-r)+1}{z-1}.
\end{align}
When the alphabet is clear from the context we will write $h_k(z)$ instead of $h_{k,r}(z)$.
\end{theorem} 

\begin{remark}
Notice that in the binary case, there is a natural correspondence between the $k$-run length system $\cS_k$ and the $(0,k-1)$-run length limited (RLL) 
constrained system $X_{k-1}$. Under the $(0,k-1)$-RLL constraint, a word $x\in X_{k-1}$ if between two ones, there are at most $k-1$ consecutive zeros. 
The correspondence is as follows. Let $w\in \cB_n(\cS_k)$ be a word of length $n$. 
Now generate $x$ from $w$ by writing $x_i=0$ if $w_i=w_{i+1}$ and $x_i=1$ if 
$w_i\neq w_{i+1}$. Notice that $x$ is a word of length $n-1$. Moreover, since $w$ contains a run of maximal length $k$, then there is a maximal sequence of 
Similarly, we can generate a word $w\in\cS_k$ from a word $x\in X_{k-1}$ by inverting the process (and deciding the the words start with $0$). 
$k-1$ zeros between two ones, so $x\in \cB_{n-1}(X_{k-1})$. Thus, the capacity is the same for both systems.

It is possible to generalize this correspondence to larger alphabets $\Sigma=[r]$ with $r>2$. 
In this case, the generalized $(0,k-1)$-RLL system $X_k$ is a system in which between every two symbols from $[r]\backslash 0$ 
there are most $k-1$ consecutive zeros. Given a word $w\in \cB_n(\cS_k)$, we can generate $x\in X_{k-1}$ by a similar rule: write $x_i=d$ for $w_{i+1}-w_i=d$ 
when addition (and subtraction) is done modulo $r$.
\end{remark}

\begin{example}
Consider the system $\cS_2$ and $\cS_3$ over the binary alphabet $\Sigma=[2]$. The capacities of these systems are 
\begin{align*}
    \ccap(\cS^{[2]}_2)&= 0.694,\\
    \ccap(\cS^{[2]}_3)&= 0.879.
\end{align*}
\end{example}

In the next lemma we will obtain some useful properties of the polynomial $h_{k,r}(z)$.
\begin{lemma}
\label{lem:poly_h}
Let $\Sigma=[r]$ with $r\geq 2$, let $h_{k,r}(z)$ be defined as in \eqref{eq:def_polynom_h}, and let $\lambda_k$ denote the maximal root of $h_{k,r}(z)$. 
Then 
\begin{enumerate}
    \item \label{lem:poly_h_1} We have $\lambda_1=r-1$ and for $k\geq 2$, $\lambda_k\in \parenv{r-\frac{1}{k}, r}$. Specifically, $\lambda_k\geq r-1$.
    \item \label{lem:poly_h_2} For every $k$, $\lambda_k\leq \lambda_{k+1}$.
    \item \label{lem:poly_h_3} For $k\geq 2$, $h_{k,r}(z)$ is increasing for $z\in [\lambda_k,\infty)$.
\end{enumerate}
\end{lemma}

\begin{IEEEproof}
The proof of the properties is straightforward.
\begin{enumerate}
    \item For $\Sigma= [r]$ with $r\geq 2$, the polynomial $h_{1,r}(z)=z-(r-1)$ and it is clear that $r-1$ is its root. 
    Assume $k\geq 2$ and plug in $z=\frac{rk-1}{k}$ to obtain  
    \[h_{k,r}\parenv{\frac{rk-1}{k}}=\frac{1-\frac{(kr-1)^k}{k^k}}{kr-k-1}.\] 
    Using the binomial formula and the inclusion-exclusion principle, we have 
    \[(kr-1)^k\geq (kr)^k-k(kr)^{k-1}.\] 
    Plugging this to the above equation we obtain 
    \[h_{k,r}\parenv{\frac{rk-1}{k}}\leq \frac{1-r^{k-1}(r-1)}{kr-k-1}\leq 0,\] 
    due to the assumption $k\geq 2$ and because $r\geq 2$. 
    
    Since $\sum_{i=0}^{k-1} (r-1)z^i= (r-1)\frac{z^k-1}{z-1}$, 
    \[h_{k,r}(r)=r^k-(r-1)\frac{r^k-1}{r-1}=1>0.\]
    Hence, there is a root in $(r-\frac{1}{k},r)$. 
    Moreover, for $k\geq 1$ we have $r-(1/k)\geq r-1$.
    
    The fact that the maximal root of $h_{k,r}$ is not greater than $r$ follows from noticing that for $z\geq r$, $h_{k,r}(z)>0$ (see \eqref{eq:def_polynom_h}).
    
\item Notice that 
    \begin{align*}
        h_{k+1,r}(z)&=z^{k+1}-(r-1)\parenv{1+z+\dots+z^k} \\
        &=(z-r)z^k+ h_{k,r}(z).
    \end{align*}
    For $0<z<r$, we have $h_{k+1,r}(z)< h_{k,r}(z)$. Together with the fact that $\lambda_{k+1}$ is the maximal root of $h_{k+1,r}(z)$, we obtain that 
    for all $z\in [\lambda_{k+1},r)$, $0\leq h_{k+1,r}(z)<h_{k,r}(z)$. Specifically, for $\lambda_{k+1}$ we have that $0<h_{k,r}(\lambda_{k+1})$ and 
    that $h_{k,r}(z)>0$ for all $z\in (\lambda_{k+1},r)$. Since $h_{k,r}(\lambda_k)=0$, we obtain that $\lambda_k<\lambda_{k+1}$.

\item First notice that for $k=1$ the polynomial $h_{1,r}$ is a constant which is non-decreasing. 
    For $k=2$, the statement is clear, so we may assume $k\geq 3$. 
    We first notice that $h_{k,r}(\lambda_k)=0$, the polynomial $h_{k,r}(z)$ is continuous for $z>r-1$ and that $h_{k,r}(r)=1>0$. 
    Together with the maximality of $\lambda_k$ we obtain that $\frac{\mathsf{d}}{\mathsf{d}z} h_{k,r}(\lambda_k)\geq 0$. 
    In fact, consider the numerator of $h_{k,r}(z)$ and notice that 
    \[\frac{\mathsf{d}}{\mathsf{d}z} z^k(z-r)+1=z^k+k z^{k-1}(z-r),\] 
    which, in turn, implies that the numerator of $h_{k,r}(z)$ has a single real minimum point at $z=\frac{rk}{k+1}$. 
    From Part \ref{lem:poly_h_1} together with $k\geq 2$, we have $\lambda_k>r-\frac{1}{k}$ which implies that the minimum point $\frac{rk}{1+k}<\lambda_k$. 
    Thus, for $z\geq \lambda_k$, it suffices to show that 
    \[\frac{\mathsf{d}}{\mathsf{d}z} z^k(z-r)+1 \geq 1,\] 
    to conclude that $h_{k,r}(z)$ is increasing for $z\geq \lambda_k$. 
    To that end, since $\lambda_k\geq r-\frac{1}{k}$, we have that for $z\geq \lambda_k$, 
    \[\frac{\mathsf{d}}{\mathsf{d}z} z^k(z-r)+1=z^k+k z^{k-1}(z-r)\geq z^k-z^{k-1}.\] 
    Notice that $\frac{\mathsf{d}}{\mathsf{d}k} (z^k-z^{k-1})=(z-1)z^{k-1}\log z$, which is positive for $z>1$. 
    Therefore, it suffices to show that $z^3-z^{3-1}>1$ for $z\geq \lambda_k$. 
    Since $z^3-z^2=1$ for $z\approx 1.47<1.5$ we obtain the desired result.
\end{enumerate}
\end{IEEEproof}

\begin{remark}
The last part of the previous lemma provides a lower bound on the capacity of the $k$-run length system $\ccap(\cS^{[r]}_k)\geq r-\frac{1}{k}$.
\end{remark}

We now turn to the analysis of the Perron eigenvectors of $A_k^{[r]}$. Our end-goal is to describe the stationary distribution of the Markov 
chain $\mu$ that corresponds to $\cS_k$. Specifically, we would like to estimate the transition probabilities between states.
We start by providing an inductive algorithm for the right Perron eigenvector of $A_k^{\Sigma}$ as a function of the right Perron eigenvector of $A_{k-1}^{\Sigma}$. Before stating the algorithm, more notations are in order. 

For $k=1$ we define $g_{1,r}(z)=\frac{1}{r-1}$ and for $2\leq k\in\N$, let us denote by $g_{k,r}(z)$ the polynomial 
\begin{align}
    \label{eq:poly_gn}
    g_{k,r}(z)=\frac{1}{r-1}z^{k-1}-\sum_{i=1}^{k-2}z^i=\frac{1}{r-1}z\parenv{h_{k-2}(z)}.
\end{align}
When that alphabet $[r]$ is clear from the context, we will write $g_k(z)$ instead of $g_{k,r}(z)$. 

\begin{lemma}
\label{lem:properties_of_gn}
Fix an alphabet $\Sigma=[r]$ with $r\geq 2$, fix $k$, let $\lambda_k$ be the Perron eigenvalue of $A_k^{[r]}$, and let $g_k(z)=g_{k,r}(z)$ be the 
polynomial defined in \eqref{eq:poly_gn}. Then 
\begin{enumerate}
    \item \label{lem:pro_gn_1} For $z\leq r$ we have $g_k(z)\leq g_{k-1}(z)\leq \dots\leq g_2(z)$.
    \item \label{lem:pro_gn_2} $g_k(\lambda_k)\geq 1$. 
    \item \label{lem:pro_gn_3} For $k\geq 2$, $g_k(\lambda_{k-2})=0$.
    \item \label{lem:pro_gn_4} For every $k\geq 1$ and $m\geq k$, $g_k(\lambda_m)\leq g_k(\lambda_{m+1})$.
\end{enumerate}
\end{lemma}

\begin{IEEEproof}
The proof of Lemma \ref{lem:properties_of_gn} is straightforward.
\begin{enumerate}
    \item Notice that for $k\geq 2$, 
\[g_{k+1}(z)=\frac{1}{r-1}z^k-\sum_{i=1}^{k-1}z^i=\frac{z-r+1}{r-1}z^{k-1}-\sum_{i=1}^{k-2}z^i=\frac{z-r}{r-1}z^{k-1}+g_k(z).\]
Therefore, for $z\leq r$ we have $g_{k+1}(z)\leq g_k(z)$ which finishes the proof.
\item Notice that
\[g_k(z)=\frac{1}{r-1}z^{k-1}-\sum_{i=1}^{k-2}z^i-1+1=\frac{1}{r-1}h_{k-1}(z)+1\geq \frac{1}{r-1}h_k(z)+1\] 
where the last inequality follows from the proof of Lemma Lemma \ref{lem:poly_h}.\ref{lem:poly_h_2}. This, in turn, implies that $g_k(\lambda_k)\geq 1$. 
\item The fact that $g_k(\lambda_{k-2})=0$ follows readily from \eqref{eq:poly_gn} and since $h_{k-2}(\lambda_{k-2})=0$.
\item The last part is clearly true for $k=1$. Moreover, since $g_2(z)=\frac{z}{r-1}$ is an increasing function of $z$ and since 
$\lambda_m\leq \lambda_{m+1}$, the statement is true for $k=2$ as well. Thus, we may assume $k\geq 3$.
Differentiate $g_k(z)$ with respect to $z$ to obtain 
\begin{align}
\label{eq:dif_g}
    \frac{\mathsf{d}}{\mathsf{d}z}g_k(z)&= \frac{1}{r-1}\parenv{h_{k-2}(z)+z\frac{\mathsf{d}}{\mathsf{d}z}h_{k-2}(z)}.
\end{align}
Since $\lambda_{k-2}$ is the maximal root of $h_{k-2}(z)$, noticing that $h_{k-2}(r)=1$ implies that $h_{k-2}(z)>0$ for $z>\lambda_{k-2}$. 
Use Lemma \ref{lem:poly_h}.\ref{lem:poly_h_2} to obtain $h_{k-2}(\lambda_m)>0$. 
From Lemma \ref{lem:poly_h}.\ref{lem:poly_h_3} we obtain that $z\frac{\mathsf{d}}{\mathsf{d}z}h_{k-2}(z)\geq 0$ for 
$z\geq \lambda_{k-2}$ which finishes the proof.
\end{enumerate}
\end{IEEEproof}

We start to work our way towards a characterization of the Markov distribution that maximizes the capacity of $\cS_k$. 
The idea is to present an algorithm that given the right Perron eigenvector for $A^{[r]}_k$, generates the right Perron eigenvalue of $A^{[r]}_{k+1}$. 
In this way, knowing the right Perron eigenvector of $A^{[r]}_1$ will make it possible to study the eigenvectors of $A^{[r]}_k$ for general $k$. 
Since for a state $i$, the transition probabilities $Q_{i,j}$ of the Markov measure that generates the sequences to be synthesized depends only on the 
$\lambda_k$ and the values of the right eigenvector, the optimal reference sequence can be determined.

The following lemma characterizes the right Perron eigenvectors of $A^{[r]}_1$, i.e., the Perron eigenvectors of the adjacency matrix of the graph that presents $\cS_1$ - the system in which every two consecutive symbols are different. 
\begin{lemma}
\label{lem:base_of_induction}
Let $x\in \R^r$ be given by $x=\1=\parenv{1,1,\dots,1}$.
Then $x$ is a right Perron eigenvector of $A^{[r]}_1$.
\end{lemma}

\begin{IEEEproof}
The lemma follows instantly from the fact that $A_1^{[r]}$ is an $r\times r$ matrix that has $0$ on its diagonal and $1$ in every other entry.
\end{IEEEproof}

For a vector $x\in \R^n$, $x=(x_0,\dots,x_{n-1})$, a set $B\subseteq [n]$, and a number $t$, we denote $x|_{B,t}$ the vector $x$ in which every coordinate 
that appears in $B$ is replaced with $t$, i.e., 
\[(x|_{B,t})_i=\begin{cases}
t, & \text{ if } i\in B, \\
x_i, & \text{ otherwise}.
\end{cases}\]
For example, if $x=(x_0,x_1,x_2,x_3,x_4,x_5,x_6,x_7)\in\R^8$, $B=\mathset{3,5}$, and $t=99$, then $x|_{B,t}=(x_0,x_1,x_2,99,x_4,99,x_6,x_7)$. 

\begin{construction}
\label{con:Perron_eig} 
Fix $1\leq k\in\N$, let $A_k^{[r]}$ denote the adjacency matrix of the graph $G([r],k)$ that presents $\cS_k$, and let $\lambda_k$ denote its 
Perron eigenvalue. 
Let $x\in \R^{r^k}$ be a right Perron eigenvector of $A_k$. 
Assume that there are $f_0,\dots,f_{r^k-1}:\R\to\R$ such that 
\[x=\parenv{f_0(\lambda_k),\dots, f_{r^k-1}(\lambda_k)}.\]
Let 
\[\tilde{x}=\parenv{f_0(\lambda_{k+1}),\dots, f_{r^k-1}(\lambda_{k+1})}.\]
For $i\in [r]$, let $B_i$ be the set of coordinates $B_i=\mathset{j\frac{r^k -1}{r-1}: j\in [r]\setminus \{i\}}$.
Let $y\in\R^{r^{k+1}}$ be the vector obtained by  
\begin{align}
\label{eq:y}
    y= \parenv{\tilde{x}|_{B_0,g_{k+1}(\lambda_{k+1})}, \tilde{x}|_{B_1,g_{k+1}(\lambda_{k+1})}, \dots, \tilde{x}|_{B_{r-1},g_{k+1}(\lambda_{k+1})}}.
\end{align} 
We will also use the notation $T(x)$ to denote the application of the construction on $x$, 
so applying the construction $k$ times will be denoted $T^k(x)$.
\end{construction}

Notice that $\frac{r^k -1}{r-1}=\sum_{l\in [k]} r^l$ and therefore the sets $B_i$ indeed contain non-negative integers. 

\begin{remark}
\label{rem:changed_entries_are_ones} 
The construction implies that in the process of constructing $T(x)$ from $x$, the entries that are replaces by $g_{k+1}(\lambda_{k+1})$ 
contain $1$ before the replacement takes place.
\end{remark}

\begin{example}
\label{ex:per_eigenvector_ex}
Consider the systems $\cS^{[2]}_2,\cS^{[2]}_3$ given in Example \ref{ex:adj_mat}. 
It is straightforward to show that the right Perron eigenvectors of $A_2^{[2]}$ is $x=(1,\lambda_2,\lambda_2,1)$ where $\lambda_2$ 
is the Perron eigenvalue of $A^{[2]}_2$ (and the largest real root of $h_{2,2}(z)$, and hence, $\ccap(\cS_2^{[2]})=\log \lambda_2$). 
A right Perron eigenvector of $A_3^{[2]}$ is $y=(1,\lambda_3,\lambda_3,\lambda_3^2-\lambda_3,\lambda_3^2-\lambda_3,\lambda_3,\lambda_3,1)$ 
where $\lambda_3$ is the Perron eigenvalue of $A^{[2]}_3$. 

It is possible to obtain $x$ from the eigenvector of $A_1^{[2]}$ using Lemma \ref{lem:base_of_induction}.
Indeed, from Lemma \ref{lem:base_of_induction}, a right eigenvector for $A_1^{[2]}$ is $(1,1)^T$. 
Notice that $x=(f_0(\lambda_1),f_1(\lambda_1))$ where $f_0=f_1=1$. Denote by $\lambda_2$ the Perron eigenvalue of $A_2^{[2]}$ 
and calculate $B_0=\mathset{1}, B_1=\mathset{0}$. Since $g_{2,2}(\lambda_2)=\lambda_2$ we obtain that $x=T((1,1))=(1,\lambda_2,\lambda_2,1)$. 

We now use $x$ to find the right Perron eigenvector for $k+1=3$. 
First, we find $f_i,\; i\in [r^k]=[2^2]$. It is evident that $f_0=f_3=1$ and $f_1=f_2=id$. Thus, 
\[\tilde{x}=(f_0(\lambda_3),f_1(\lambda_3),f_2(\lambda_3),f_3(\lambda_3))=(1,\lambda_3,\lambda_3,1),\] 
where $\lambda_3$ is the Perron eigenvalue of $A_3^{[2]}$ or alternatively, the maximal root of 
\[h_{3,2}(z)=z^3-z^2-z-1.\] 
For $i\in [2]$, we calculate $B_i$ and obtain that $B_0=\mathset{1\cdot\frac{r^k-1}{r-1}}=\mathset{3}$ and $B_1=\mathset{0}$. 
Next, we find $g_3(\lambda_3)= \frac{\lambda_3^2}{2}-\lambda_3$. 
Overall, we have 
\[\tilde{x}|_{B_0,g_3(\lambda_3)}=\tilde{x}|_{\mathset{3},\lambda_3^2-\lambda_3}= (1,\lambda_3,\lambda_3,\lambda_3^2-\lambda_3),\]
and 
\[\tilde{x}|_{B_1,g_3(\lambda_3)}=\tilde{x}|_{\mathset{0},\lambda_3^2-\lambda_3}= (\lambda_3^2-\lambda_3,\lambda_3,\lambda_3,1) .\]
Combining the above we obtain that 
\[y=T(x)=(1,\lambda_3,\lambda_3,\lambda_3^2-\lambda_3,\lambda_3^2-\lambda_3,\lambda_3,\lambda_3,1).\]
\end{example}

The following lemmas provide some insight on the structure of the Perron eigenvector.
\begin{lemma}
\label{lem:all_pos}
Let $\Sigma=[r]$ and let $\1\in \R^r$ be the vector of all ones (the eigenvector of $A^{[r]}_1$). 
For every $k\in\N$, let $y=T^k(\1)\in\R^{r^k}$ be the vector obtained after applying the construction in \eqref{eq:y} on $\1$ for $k$ times. 
Then all the entries of $y$ are strictly positive.
\end{lemma}

\begin{IEEEproof}
In order to prove the statement for $k\geq 1$, we notice that it is sufficient to show that $g_m(\lambda_k)>0$ for all $m\leq k$. 
Indeed, from the construction in \eqref{eq:y}, all the arguments that appear in $y$ have the form $g_m(\lambda_k)$ for $m\leq k$.
Now the lemma follows immediately from parts \ref{lem:pro_gn_2} and \ref{lem:pro_gn_4} of Lemma \ref{lem:properties_of_gn}. 
\end{IEEEproof}

\begin{lemma}
\label{lem:pos_of_ones}
Fix $\Sigma=[r]$. For every $k\in \N$, let $y=T^k(\1)\in\R^{r^k}$ be the vector obtained by applying \eqref{eq:y} $k$ times on $\1$. 
Then $y$ contain ones in positions $\mathset{j\frac{r^k-1}{r-1} : j\in [r]}$.
\end{lemma}

\begin{remark}
Notice that Lemma~\ref{lem:pos_of_ones} suggests that if we enumerate the positions of the vector $y$ as base-$r$ numbers then $y$ contains ones in positions of the form 
$\underbrace{jj\dots j}_{k\text{ times }}$, for $j\in [r]$.
\end{remark}

\begin{IEEEproof}
We prove the lemma using induction. For $k=1$ we have that the all ones vector contain ones in all the positions, i.e., in positions $[r]$. Assume that the lemma is correct for $k-1$ and we will show it is correct for $k$. Let $x\in \R^{r^{k-1}}$ be the vector obtained after applying $T$ for $(k-1)$ times. From the induction hypothesis, $x$ contains ones only in positions 
$\mathset{j\frac{r^k-1}{r-1} : j\in [r]}$. Thus, $\tilde{x}$ contain ones at the exact same positions. 
From the definition of $B_i$ we obtain that $\tilde{x}|_{B_i,g_{k+1}(\lambda_{k+1})}$ contains ones only in position $i\frac{r^{k-1}-1}{r-1}$. 
This implies that $y$ contains ones in positions $\mathset{i r^{k-1}+i\frac{r^{k-1}-1}{r-1} : i\in [r]}=\mathset{i\frac{r^k-1}{r-1} : i\in [r]}$, and the proof follows.
\end{IEEEproof}

We next claim that $T(x)$ is indeed a Perron eigenvector. 
\begin{claim}
\label{cl:right_eigen}
For an alphabet $[r]$ and $k\geq 1$. 
Let $\1$ be the all one vector comprises of $r$ ones (the right Perron eigenvector of $A^{[r]}_1$), then for $k\geq 1$, $y=T^{k-1}(\1)$ is a right Perron eigenvector of $A_k$.
\end{claim}

The proof of Claim \ref{cl:right_eigen} follows immediately from the following lemma. 
\begin{lemma}
\label{lem:diff_pers}
Fix alphabet $\Sigma=[r]$ and let $\1_r$ be the right Perron eigenvector of $A^{[r]}_1$. 
For $k\geq 1$, the vector $y=T^{k-1}(\1)\in \R^{r^k}$ has the following form. 
Let $i=(i_{k-1}\dots i_0)\in [r^k]$ be considered in its base-$r$ presentation and let $l_i\in [k]$ be the largest number for which $i_0=i_1=\dots=i_{l-1}$, 
i.e., $l_i$ denotes the number of repeated least significant symbols in the base-$r$ representation of $i$. 
Then 
\[y_i=\begin{cases}
1& \text{if } l_i=k\\
g_{l_i+1}(\lambda_k) & \text{if } l_i<k,
\end{cases}\]
where $\lambda_k$ is the Perron eigenvalue of $A^{[r]}_k$.
\end{lemma}

\begin{example}
As seen in Example \ref{ex:per_eigenvector_ex}, for $k=3$, a right Perron eigenvector of $A_3^{[2]}$ is $y=T^2(\1)=(1,\lambda_3,\lambda_3,\lambda_3^2-\lambda_3,\lambda_3^2-\lambda_3,\lambda_3,\lambda_3,1)$ 
where $\lambda_3$ is the Perron eigenvalue of $A^{[2]}_3$. 
Indeed, in positions $i=0,7$ (in binary representation - $000,111$), there is $1$. 
In positions $i=1,2,4,6$ (positions $001,010,101,110$), the value of $l_i=1$ and so $y_i=g_2(\lambda_k)=\lambda_3$, 
and in positions $i=3,5$ (positions $011,100$), we have $l_i=2$ and so $y_i=g_3(\lambda_k)=\lambda_3^2-\lambda_3$. 
\end{example}

\begin{IEEEproof}[Proof of Lemma~\ref{lem:diff_pers}]
The proof follows by induction. For the base of induction we note that $T(\1)$ contains $1$ in positions $00,11,\dots, (r-1)(r-1)$ and contains 
$\frac{\lambda_2}{r-1}$ in the rest of the positions, where $\lambda_2$ is the Perron eigenvalue of $A^{[r]}_2$. 
Now assume this is true for $k-1$ and we show it holds for $k$. Let $x=T^{k-1}(\1)$ and let $y=T(x)$. 
By Construction \ref{con:Perron_eig}, the positions in $y$ that contain $g_{k+1}(\lambda_{k+1})$ are positions such that written in their base-$r$ 
representation have the form $ab^k$ for $a\neq b \in [r]$. According to Lemma \ref{lem:pos_of_ones}, positions of the form $a^{k+1}$ contain $1$. 
Position $i=(i_{k-1}\dots i_0)$ in $x$ with $l_i<k$ correspond to positions $ai=(a i_{k-1}\dots i_0)$ with $a\in [r]$ in $y$, 
and will remain with the same value $l_{ai}=l_i$. 
\end{IEEEproof}

\begin{IEEEproof}[Proof of Claim \ref{cl:right_eigen}]
The proof now follows from a straightforward calculation of $A^{[r]}_k y$ where $y=T^{k-1}(\1_r)$ and $\1_r$ is the vector comprises of $r$ ones. 
This can be done using Lemma \ref{lem:diff_pers} and Lemma \ref{lem:adj_mat_of_S_k}. 
Let $i=(i_{k-1}\dots i_0)\in [r^k]$ and consider $(A_k y)_i$. Let us denote by $l_i$ the number of repeated least significant symbol in $i$. 
Lemma \ref{lem:diff_pers} implies that $y_i=g_{l_i+1}(\lambda_k)$.
\begin{enumerate}
    \item Case $1$: $l_i\leq k-2$. 
    Using Lemma \ref{lem:adj_mat_of_S_k} we obtain that 
    \[(A_k y)_i=\sum_{j=0}^{r-1} y_{i_{k-2}\dots i_0 j}.\]
    Since $l_i\leq k-2$ the sum $(A_k y)_i$ comprises of $r-1$ values $\frac{\lambda_k}{r-1}$ and a single value 
    $g_{l_i+2}(\lambda_k)$. 
    Since 
    \[g_{l_i+2}(\lambda_k)+\lambda_k=\frac{1}{r-1}\lambda_k^{l_i+1}-\sum_{i=1}^{l_i}\lambda_k^i +\lambda_k=\lambda_k \parenv{g_{l_i+1}(\lambda_k)},\] 
    we have 
    \[(A_k y)_i=\lambda_k g_{l_i+1}(\lambda_k).\]
    
    \item Case $2$: $l_i=k-1$. 
    Using Lemma \ref{lem:adj_mat_of_S_k} we obtain that 
    \[(A_k y)_i=\sum_{j=0}^{r-1} y_{i_{k-2}\dots i_0 j}.\]
    In this case we have that the sum $(A_k y)_i$ comprises of $r-1$ values $\frac{\lambda_k}{r-1}$ and a single value $1$. 
    Notice that 
    \[\lambda_k g_k(\lambda_k)=\frac{1}{r-1}\lambda_k^k-\sum_{i=2}^{k-1}\lambda_k^i=h_k(\lambda_k)+\lambda_k+1=\lambda_k+1.\] 
    Therefore, in this case as well, 
    \[(A_k y)_i=\lambda_k+1=\lambda_k g_k(\lambda_k).\]
    
    \item Case $3$: $l_i=k$. 
    Let $i=(a\dots a)$ for $a\in [r]$.
    Using Lemma \ref{lem:adj_mat_of_S_k} we obtain that 
    \[(A_k y)_i=\sum_{j\in [r]\backslash \{a\}} y_{a\dots a j}.\]
    In this case we have that the sum $(A_k y)_i$ comprises of $r-1$ values $\frac{\lambda_k}{r-1}$ and $y_i=1$. Thus, we obtain 
    \[(A_k y)_i=\lambda_k.\]
\end{enumerate}
In all three cases above we obtain that $(A_k y)_i=\lambda_k y_i$. Lemma \ref{lem:all_pos} together with Perron-Frobenius theorem imply that $y$ is 
the right Perron eigenvector of $A_k$, as claimed.
\end{IEEEproof}

\begin{claim}
\label{cl:last}
Let $\cS_k$ be the $k$-run length constrained system over the alphabet $\Sigma=[r]$ and let $\mu$ be the Markov measure that corresponds to $\cS_k$ 
with $\pi$ its stationary distribution and $Q$ its transition matrix. 
For $i\in [r^k]$ we denote by $l_i$ the number of repeated least significant symbol. 
Then for every $i,j,t\in [r^k]$, if $l_j<l_t$ and there are edges $(i,j),(i,t)$, then $Q_{i,j}>Q_{i,t}$. 
\end{claim}

\begin{IEEEproof}
The proof follows immediately since $Q_{i,j}=\frac{y_j}{\lambda_k y_i},Q_{i,t}=\frac{y_t}{\lambda_ky_i}$ where $y$ is the right Perron eigenvector, since 
$y_t=g_{l_t+1}(z),y_j=g_{l_j+1}(z)$ and from Lemma \ref{lem:properties_of_gn}.\ref{lem:pro_gn_1}.
\end{IEEEproof} 

Finally, we prove Theorem \ref{th:main2}.
\begin{IEEEproof}[Proof of Theorem \ref{th:main2}]
The proof follows from Claim \ref{cl:last} since the claim suggests that the probability of repeating the last symbol is the smallest. 
This implies that $\mu(ab^i)$ is decreasing as a function of $i$. 
In addition, since $Q_{i,j}=\frac{y_j}{\lambda_k y_i}$, where $y$ is the right Perron eigenvector, and since by Lemma \ref{lem:diff_pers}, 
$y_i,y_j$ depend only on the numbers $l_i,l_j$ of repeated least significant symbols, $\mu(ab^i)=\mu(ba^i)=\mu(ac^i)$, for every $a,b,c\in [r]$ with $a\neq b$ and $a\neq c$. 
\end{IEEEproof}

\section{Conclusion and Outlook}\label{sec:ConcandOut}
In this paper, we studied the single batch settings, in which information sequences appear in the same set and are synthesized with 
respect to a reference $\sR$. We showed that the optimal reference sequence is $\sRS=\overline{012\dots (r-1)}$. 
Throughout the analysis, we provided an explicit formula for the right Perron eigenvector of the adjacency matrix corresponds to the 
constrained system $\cS_k$. A complete analysis of the system $\cS_k$ will be achieved if an explicit formula for the left Perron eigenvector will be 
found. This may also provide some concrete bounds on the cost of synthesis $\s{cost}(\cS)$. We leave this endeavour for a future research. 


\bibliographystyle{IEEEtranS}
\bibliography{allbib.bib}
\end{document}